# Highly efficient generation of pulsed photon pair using a bulk Periodically Poled Potassium Titanyl Phosphate

Bao-Sen Shi and Akihisa Tomita

Imai Quantum Computation and Information Project, NEC Tsukuba Laboratories,
ERATO, Japan Science and Technology Corporation (JST)
Fundamental Research Laboratories, NEC, 34 Miyukigaoka, Tsukuba, Ibaraki,
305-8501, Japan

Abstract

In this paper, we demonstrate efficient generation of collinearly propagating, degenerate pulsed photon pairs based on a bulk Periodically Poled Potassium Titanyl Phosphate pumped by an ultrashort pulse laser. Using a single-mode fiber as a spatial mode filter, we detect about 3200 coincidence counts per second per milliwatt pump power. After we consider main losses in our experiment, the inferred coincidence counts are about $1.09\times10^{5}$ per second per milliwatt pump power. This is the very promising for realization of sources for quantum communication and metrology.

Sources creating the entangled photon pairs are an essential tool for a variety of fundamental quantum mechanical and quantum information experiments. [1-7] At present, the most accessible and controllable source of entanglement is the process of Spontaneous Parametric Down-Conversion (SPDC) in a nonlinear crystal [8,9]. Although a lot of important results have been obtained, but still many experiments like teleportation, swapping suffer from the lower photon-pair production leading to low signal-to-noise ratio and long measurement times. More recently, periodically poled bulk crystals or waveguides were used to generate the photon pair, and experimentally were shown to be very efficient compared to bulk crystals [10-17]. A periodically-poled crystal, like a Periodical-Poled Lithium Niobate (PPLN) or a Periodically-poled Titanyl Phosphate (PPKTP), with an appropriate grating period, permit efficient three-wave mixing at user-selectable wavelengths by the technique of quasi-phase matching (QPM). QPM allows one to utilize larger nonlinear coefficients in some materials, thus leading to substantially higher two-photon production rates. Recent works are more focus on SPDC by cw laser. [10-12, 14-17] SPDC by pulse pump using a waveguide is considered in Ref. [13], the high ratio of coincidence to single count is shown with a multimode fiber. In this work, we have studied the generation of collinearly propagating pulsed photon-pair by pumping a type-I phase matching bulk PPKTP crystal with an ultrashort pulse laser. Using a single mode fiber as a spatial filter, we detect about 3200 coincidence counts per second per milliwatt pump power. After we consider main losses in our experiment, the inferred coincidence counts are about $1.09\times10^5 / s / mw$. To be our best knowledge, this is the highest photon coincidence count in ultrashort pulse cases. It is the very promising for the realization of sources for quantum information and metrology.

The PPKTP we use in experiment is bought from Raicol Crystal Company. The crystal is cut at type-I phase matching, with 1.05mm (z) x 2.1mm (y) x 2.12mm (x) size, where, z, y, x means height, width and length respectively. This PPKTP is antireflection coated on both facets at 800nm and 400nm. And it is placed at a temperature-stabilized LD holder with built-in Peltier device with a stability of $\pm 0.01^0 C$. The grating period of PPKTP is about 3.25 micrometer. In type-I phase matching, the SPDC photon pairs and pump laser have the same polarizations. We firstly characterize the PPKTP by performing the second harmonic generation (SHG) using a strong 800-nm femtosecond laser from a mode-locked Ti: sapphire laser (coherent: Vitesse). The pulse width of laser is less than 100fs. Repetition rate is 80MHz. Beam size is about 1.3mm. We do two experiments: one is for checking the SHG power dependent on pump power; another is for checking the SHG power dependent on the temperature of PPKTP crystal. The femtosecond pulses are focused with a 20-cm focal length lens. After SHG process, we use a fused silica prism to separate the pump laser and SHG, and further use two blue filters (color glass filter BG40) with transmission coefficient 85% at 400nm to cut the remained pump laser completely. Figure 1 shows the experimental SHG power data when we fix the temperature at $29^0 C$, and change the pump power gradationally. The spectrum of SHG is shown in Fig. 2. The measured bandwidth is about 0.18nm with 150.5mw pump power. Figure 3 shows the SHG power dependent on the temperature with fixed pump power 52mw. From experimental data, we can see that the SHG output is not very sensitive to the temperature. This is because of very wide bandwidth of pump laser (about 11nm). To change the temperature is equal to change the peak wavelength of SHG, but this turning coefficient is quite small near 390nm spectral region ($\sim 2.4\times10^{-2}\,nm/^{0}C$) [18].

For the coincidence measurement, we use the follow experimental setup, which is shown in Fig. 4. The output of the mode-locked Ti sapphire laser is doubled in a 1-mm type-I beta-barium-borate (BBO) to generate blue pulses with center wavelength 400nm. Before it pumps the PPKTP crystal, the blue pulses firstly transmit through two about 1.5mm pinholes, and are attenuated by a continuous variable attenuator. The spectrum of blue pulses measured after attenuator is about 3.2nm. The peak wavelength is 399.44nm. A lens with 20cm focus length focuses blue pulses. The power of blue pulses measured after lens before PPKTP crystal is 1mw. PPKTP is still placed at a temperature-stabilized LD holder with built-in Peltier device with a stability of $\pm 0.01^{0}C$. And the temperature of PPKTP is $8^{0}C$. Firstly, we use two red filters (color glass filter RG715) with transmission coefficient 90% at 800nm to cut the remained pump pulses, then we couple the SPDC photon pair to a 2-m single-mode fiber (P1-4224-FC-2, NA=0.12, manufactured by 3M, design wavelength 820nm, operating wavelength range is typically 50nm below and 200nm above the design wavelength.). The objective lens (NA=0.15) is placed at the position where is almost symmetric position to lens with the center of crystal. The output of the fiber is coupled to a fiber 50/50 beamsplitter with operating wavelength 798nm (manufactured by OFR) by FC-to-FC coupling. The each output of fiber beamsplitter is sent to a single-photon detector (PerkinElmer SPCM-AQR-14-FC). The outputs of detectors are sent to a coincidence circuit for coincidence counting. The coincidence circuit consists of a time-to-amplitude converter and single-channel analyzer (TAC/SCA, ORTEC 567) and a counter (SR400, Standford research systems). The time window of coincidence counting is 4ns. Firstly, we use the 800nm pulses transmitted through PPKTP to align the single-mode fiber. The measured coupling efficient is about 14.4%. The loss of FC-to-FC coupling is about 50%. Under these conditions, we detect the coincidence counts about 3200/s. The two single counts are 52100/s and 129300/s. (here all data are without accident counts) The main reason why there is large difference between two single detectors, we think is from single mode fiber. The operating wavelength of fiber is not symmetric to the center wavelength 800nm, which makes the spectrum of SPDC detected asymmetry to center wavelength. This will introduce the asymmetric splitting ratio. When we use a narrow bandfilter to instead of a redfilter, we can get the almost same single counts. If we consider the losses shown previously, and further consider the 50% loss with fiber beamsplitter, the inferred coincidence counts is about $1.09\times10^5 / s / mw$. To be the best of our knowledge, this is the highest coincidence count in ultrashort pulse case. We also measure the coincidence counts dependent on temperature of PPKTP. The experimental data are shown in Fig. 5. From data, we can find that the coincidence counts decreases when the temperature of PPKTP is increased. At the same time, single counts also decreases. The single count of one detector $D_1$ is shown in Fig.6. In our experiment, it seems that the PPKTP is more efficient in the lower temperature. We find the optimal temperature is less than $10^0 C$. We also decrease the temperature to $3^0 C$ by each step $1^0 C$, and the measured coincidence counts are almost unchanged. But at the same time, the possibility of liquidation on the surfaces of LD holder and PPKTP will increase under the condition of our laboratory. The optimal temperature is dependent on the peak wavelength of pump pulse. From experimental data, we find that the optimal temperature for SHG and SPDC is different. We think the main reason is from the difference between the center wavelengths. In SHG process, we measure the center wavelength is about 400.194nm from the spectrum at optimal temperature, but the peak wavelength of blue pulse used for SPDC is about 399.44nm. This difference introduces the different optimal temperature.

In conclusion, we have demonstrated efficient generation of pulsed photon pairs using a PPKTP crystal. We detect about 3200 coincidence counts per second per mw pump power, which infers the coincidence counts about 109000/s/mw if main losses of our experimental setup are considered. To be our best knowledge, this is the best data in the pulse case. It is the very promising for realization of sources for quantum communication and metrology.

We thank Prof. Imai for the support. We thank Dr. S. Kouno and K. Usami for many valuable discussions and technical supports.

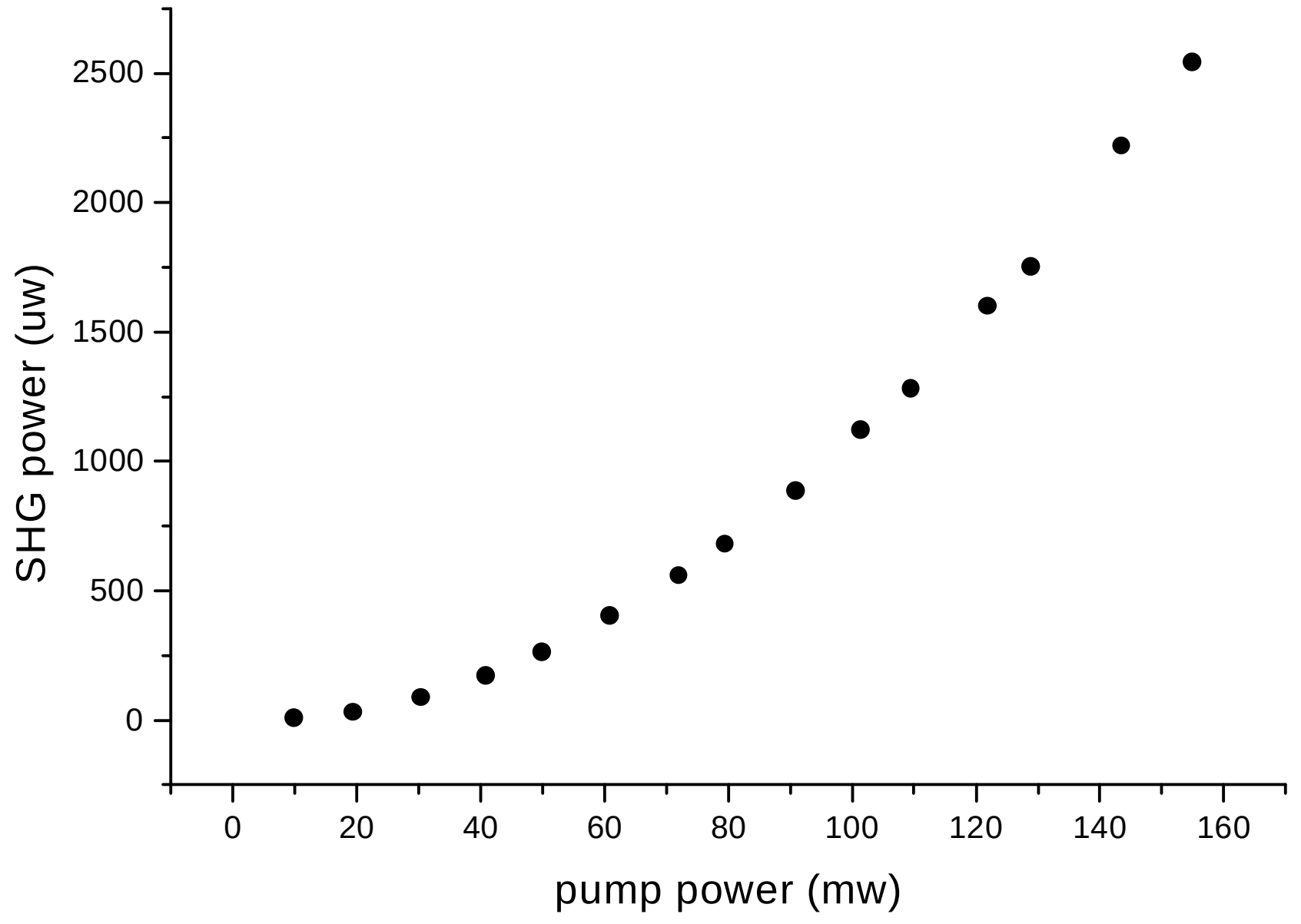

Fig.1. SHG power plotted against the power of fundamental pump laser. The temperature of PPKTP crystal is $29^0C$. All data are not corrected with the transmission rate of bluefilter.

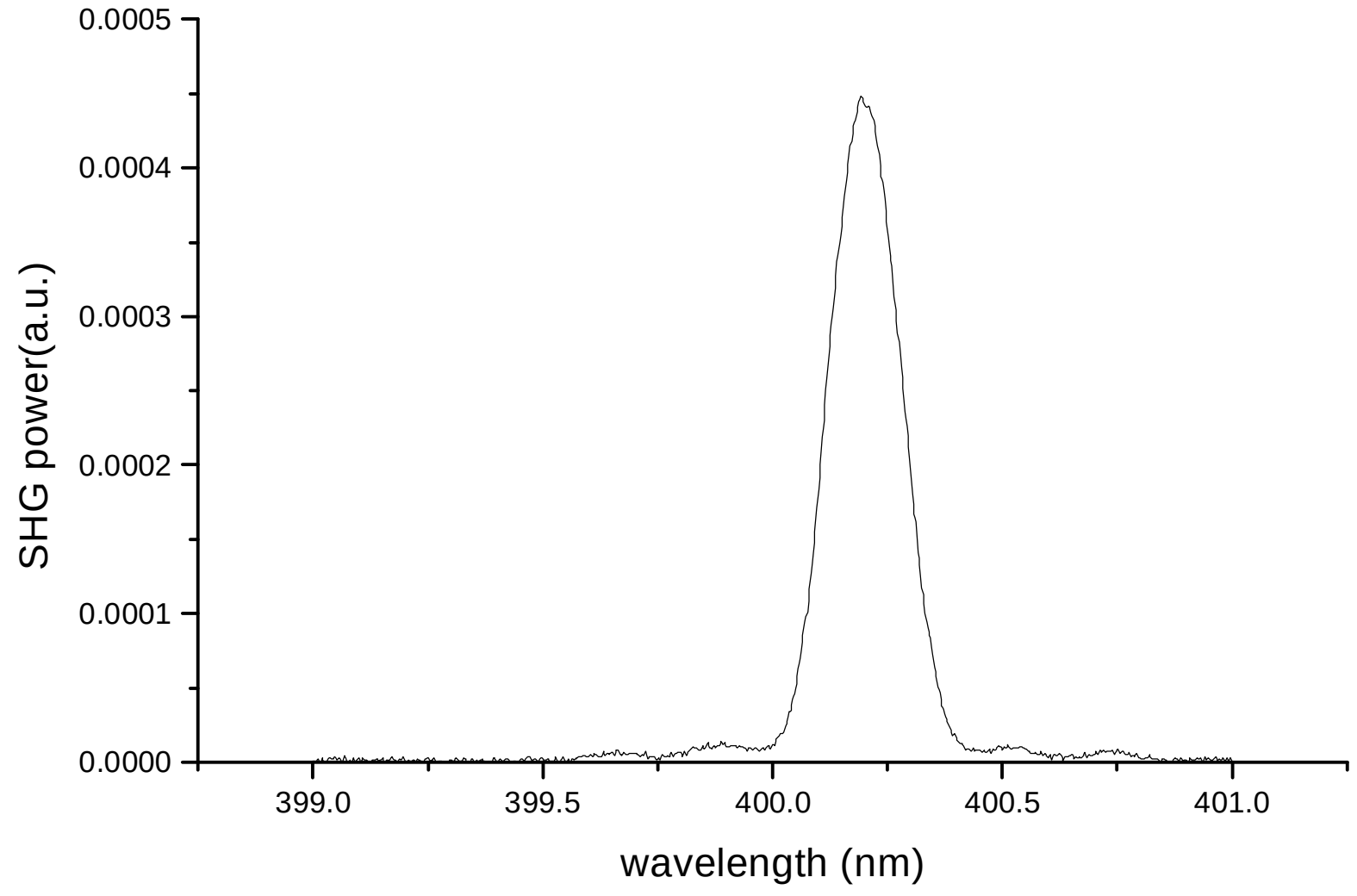

Fig. 2. Second-harmonic pulse spectrum shown on a linear scale. The FWHM is 0.18nm with peak wavelength 400.194nm. The power of pump laser is 150.5mw. Temperature of PPKTP is $29^0C$.

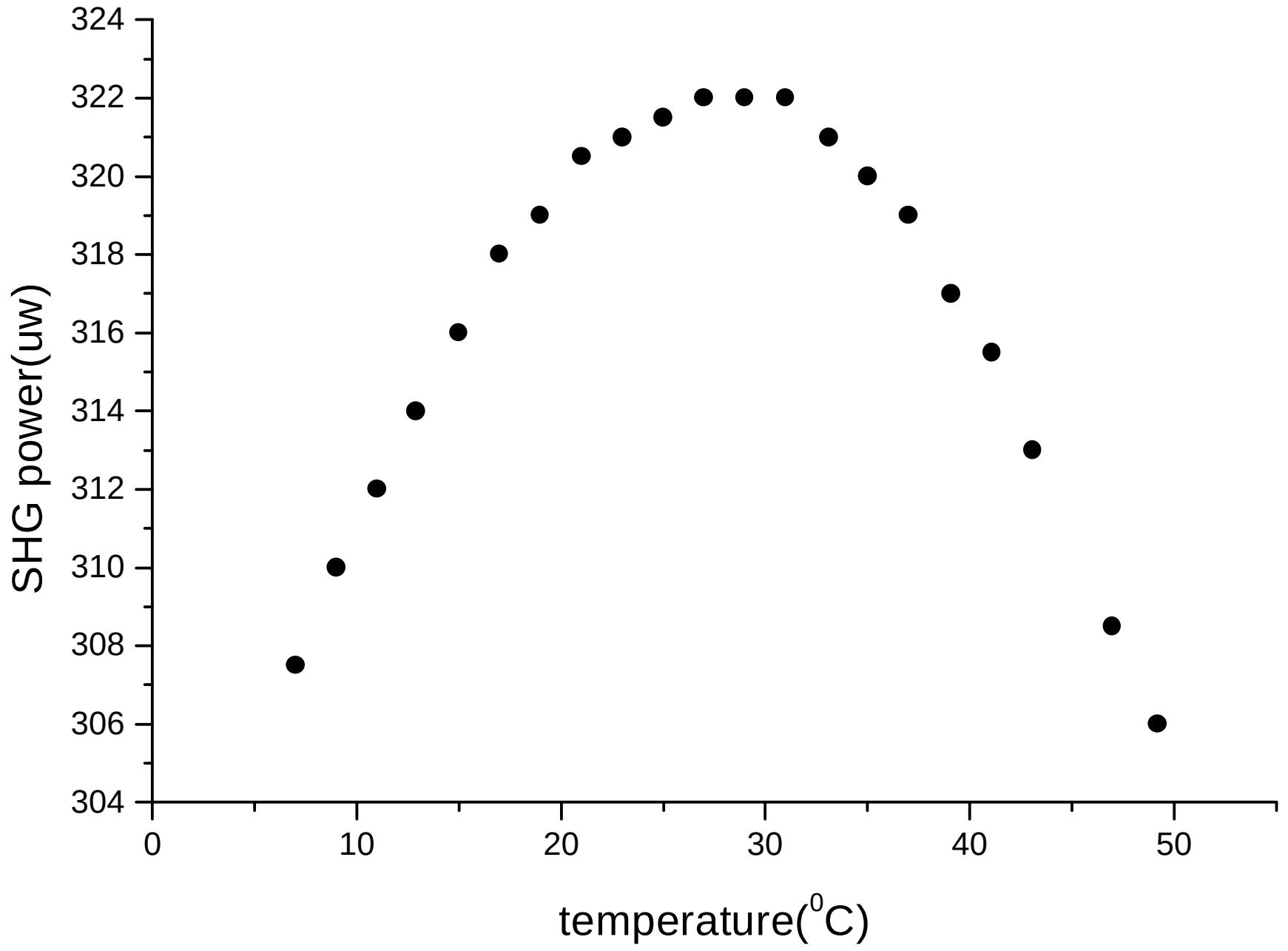

Fig.3. Second-harmonic power plotted against the temperature of PPKTP crystal. Pump power is 52mw.

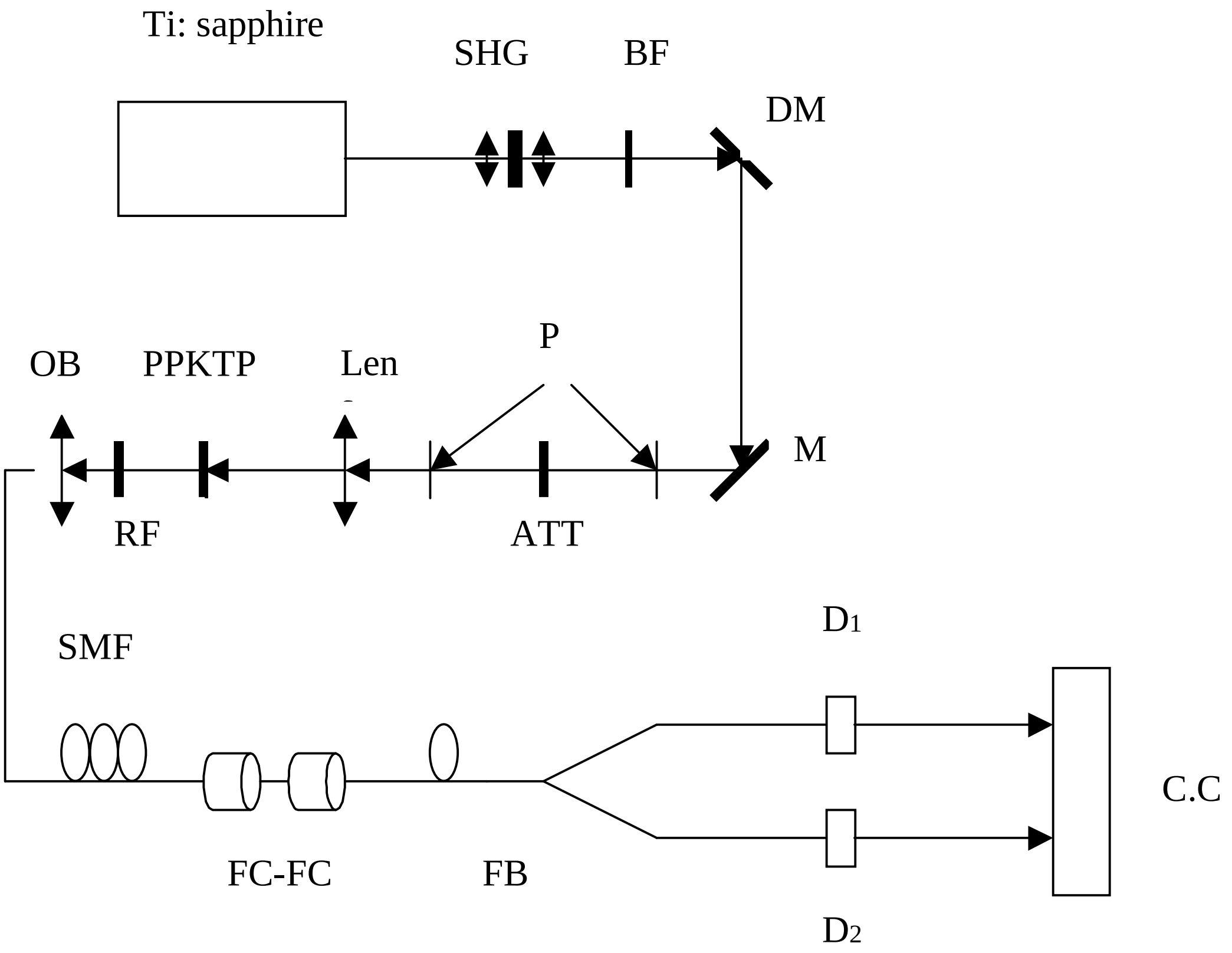

Fig. 4. Experimental setup. SHG: second harmonic generation; BF: bluefilter; DM: diachronic mirror; M: mirror; P: pinhole; ATT: attenuator; RF: redfilter; OB: objective lens; SMF: single mode fiber; FB: fiber 50/50 beamsplitter; D: detector; C.C: coincidence circuit.

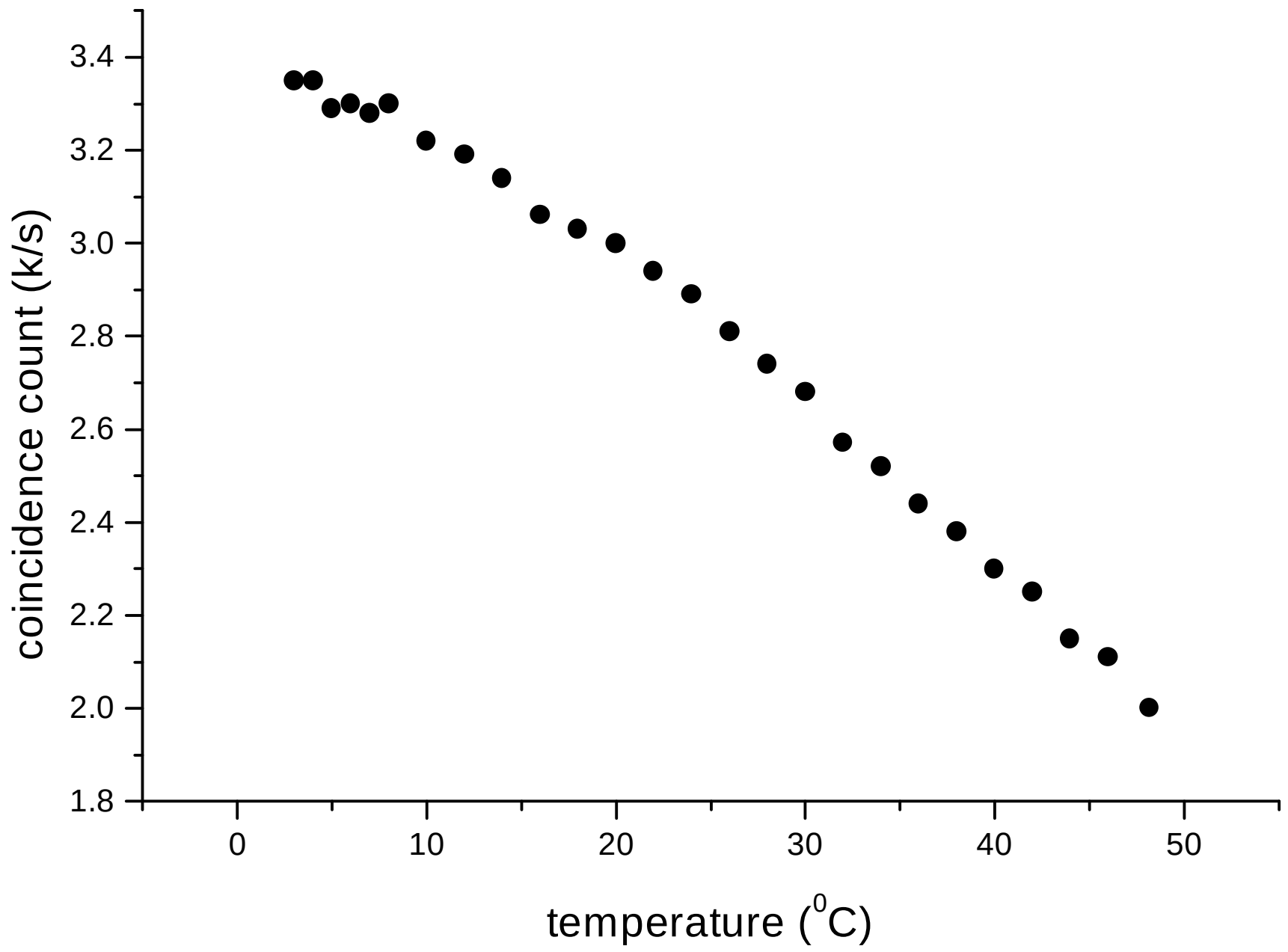

Fig.5. Coincidence counts (with accident counts) plotted against the temperature of PPKTP crystal. Power pump is 1mw. Coincidence window time is 4ns.

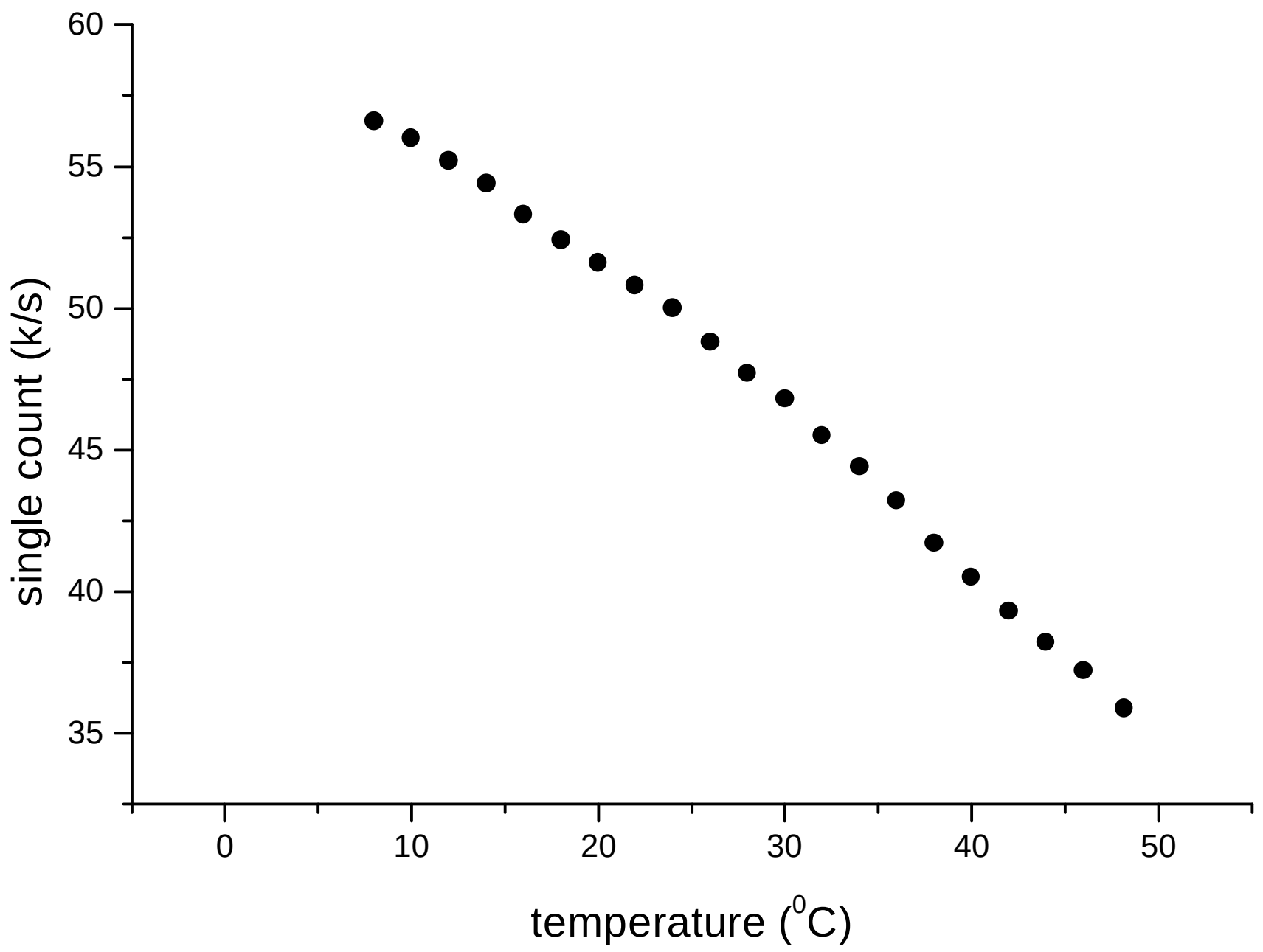

Fig.6. Single counts (with accident counts) of detector $D_1$ plotted against the temperature of PPKTP. Pump power is 1mw.